\documentclass[a4paper,11pt]{article}
\usepackage{pos}
\usepackage{float}
\usepackage{siunitx}
\usepackage{mathtools}
\usepackage{physics}
\usepackage[noabbrev, nameinlink]{cleveref}
\creflabelformat{equation}{#2\textup{#1}#3}
\usepackage{orcidlink}
\title{Enhanced Sampling Techniques for Lattice Gauge Theory}

\author*{Timo Eichhorn\textsuperscript{\orcidlink{0000-0001-9370-3642}}}
\author{Gianluca Fuwa\textsuperscript{\orcidlink{0000-0002-8195-4900}}}
\author{Christian Hoelbling\textsuperscript{\orcidlink{0000-0001-5715-1086}}}
\author{Lukas Varnhorst\textsuperscript{\orcidlink{0000-0002-3718-0143}}}

\affiliation{Department of Physics, University of Wuppertal\\
  Gaußstraße 20, D-42119 Wuppertal, Germany}

\emailAdd{timo.eichhorn@protonmail.com}

\abstract{In theories with topological sectors, such as lattice QCD and four-dimensional SU(N) gauge theories with periodic boundary conditions, conventional update algorithms suffer from topological freezing due to large action barriers separating distinct sectors. With appropriately constructed bias potentials, Metadynamics and related enhanced sampling techniques can mitigate this problem and significantly reduce the integrated autocorrelation times of the topological charge and associated observables. We test strategies to accelerate the buildup of bias potentials and the possibility of extrapolating potentials from small to large volumes. We also investigate the effectiveness of orthogonal algorithmic improvements, such as longer HMC trajectories and HMC variants, which may benefit conventional simulations as well.}

\FullConference{The 42nd International Symposium on Lattice Field Theory (LATTICE2025)\\
2-8 November 2025\\
Tata Institute of Fundamental Research, Mumbai, India}

\begin{document}
\maketitle

\section{Introduction}
\label{sec:Introduction}
Markov chain Monte Carlo simulations of lattice field theories suffer from critical slowing down as the continuum limit is approached. In topologically non-trivial theories, this includes a particularly severe form of critical slowing down known as topological freezing \cite{Alles:1996vn, Orginos:2001xa, DeGrand:2002vu, Noaki:2002ai, Aoki:2005ga, Schaefer:2009xx, Schaefer:2010hu}. To address these issues, the search for more efficient sampling algorithms has received considerable attention in recent years, including multiple related contributions that were presented at this conference \cite{Kanwar:2025wuc, Bonanno:2026tle, Matsumoto:2026puv, Singh:2026zls}. For more detailed overviews, we refer to recent review articles \cite{Finkenrath:2023sjg, Kanwar:2024ujc, Boyle:2024nlh, Finkenrath:2024ptc}.

Many of the proposed methods are based on introducing an auxiliary distribution, or a sequence of such distributions, that is easier to sample than the original one, together with a mechanism for recovering expectation values with respect to the original physical distribution. In the most extreme case, the auxiliary distribution can be chosen to be (nearly) trivial, as in some approaches that are based on trivializing maps \cite{Luscher:2009eq}, normalizing flows \cite{Albergo:2019eim}, or diffusion models \cite{Wang:2023exq}. In other cases, the auxiliary distribution remains closer to the physical theory, as in parallel tempering on boundary conditions \cite{Hasenbusch:2017unr, Bonanno:2024zyn} and most other parallel tempering-based methods. Generally, there is a trade-off between the simplicity of sampling from the auxiliary distribution and the difficulty of relating it to the original physical distribution. This trade-off depends not only on the distributions themselves, but also on the observables of interest.

The approach we proposed \cite{Eichhorn:2023uge} combines ideas from enhanced sampling methods based on collective variables (CVs) with parallel tempering. Specifically, we introduce at least one auxiliary stream with a CV-dependent bias potential and couple it to an unbiased measurement stream within a parallel tempering setup. This enhances transitions between topological sectors without requiring any reweighting. With a suitably constructed bias potential, the barriers between topological sectors can be effectively flattened, while swap acceptance rates remain reasonably high because the auxiliary and physical distributions retain sufficient overlap by design. As demonstrated in a proof-of-concept study for $N_f = 2$ staggered simulations at unphysical pion masses \cite{Eichhorn:2025aph}, there are no fundamental obstacles to applying this method to simulations with dynamical fermions. We are currently extending this approach to more realistic parameters, including the computation of the topological susceptibility at $T > T_c$ for fine lattice spacings in the range from \SI{0.05}{\femto\m} to \SI{0.02}{\femto\m}.

In this work, we focus on accelerating the construction of a suitable bias potential through an improved buildup strategy and extrapolation to larger volumes. In addition, we consider orthogonal algorithmic improvements to the Hybrid Monte Carlo (HMC) algorithm \cite{Duane:1987de} itself.
\section{Collective variable bias potential simulations}
\label{sec:Collective_variable_bias_potential_simulations}
The general idea of CV-based bias potential simulations is to introduce an additional potential $V(\mathbf{s}(U))$ to the system. This potential depends on a set of collective variables $s_i(U)$, which are in principle arbitrary, but typically differentiable functions of the fundamental degrees of freedom $U$. We can then distinguish between the original physical probability distribution $p(U) \propto e^{-S(U)}$, and the modified, biased probability distribution $p_{V}(U) \propto e^{-S(U) -V(\mathbf{s}(U))}$. Additionally, we may want to sample from a prescribed target distribution\footnote{We use the term target distribution to refer to a desired marginal distribution in the space of CVs, not the physical distribution itself, as is common in other contexts.} $p_{\mathrm{tg}}(\mathbf{s})$, e.g., a uniform distribution or a well-tempered distribution $p_{\mathrm{tg}}(\mathbf{s}) \propto p(\mathbf{s})^{1 / \gamma}$ (with $\gamma \geq 1$). From this perspective, the bias potential required to realize a given target distribution is, up to an irrelevant additive constant, given by
\begin{equation}
    V_{\mathrm{tg}}(\mathbf{s}) = \ln\biggl(\frac{p(\mathbf{s})}{p_{\mathrm{tg}}(\mathbf{s})}\biggr).
\end{equation}
The success of these methods depends crucially on the quality of the CVs, since they must capture the relevant action barriers in configuration space. In the context of topological freezing, a non-integer definition of the topological charge is a natural choice. In the following, we always use the clover-based definition after four stout smearing steps \cite{Morningstar:2003gk} with smearing parameter $\rho = 0.12$. Further improvements may be possible, e.g., via constructions of CVs with enhanced overlap with the problematic slow modes, or through different smearing kernel actions \cite{GarciaPerez:1993lic, deForcrand:1997esx, Tanizaki:2024zsu, Butti:2025rnu}.

\subsection{Variationally enhanced sampling}
\label{subsec:Variationally_enhanced_sampling}
In contrast to Metadynamics \cite{Laio_2002}, where the bias potential is constructed as a sum of Gaussians, variationally enhanced sampling (VES) \cite{PhysRevLett.113.090601} treats the bias potential as a parametrized function $V(\mathbf{s}; \boldsymbol{\alpha})$ of the collective variables $\mathbf{s}$ and determines the parameters $\boldsymbol{\alpha}$ variationally. Given a target distribution $p_{\mathrm{tg}}(\mathbf{s})$, one introduces the convex functional
\begin{equation}
    \Omega[V] = \ln( \frac{\int \mathrm{d}\mathbf{s}\, p(\mathbf{s})\, e^{- V(\mathbf{s})}}{\int \mathrm{d}\mathbf{s}\, p(\mathbf{s})} ) + \int \mathrm{d}\mathbf{s}\, p_{\mathrm{tg}}(\mathbf{s})\, V(\mathbf{s}).
\end{equation}
The parameters $\boldsymbol{\alpha}$ are then varied to minimize $\Omega[V]$, which is equivalent to minimizing the expression $D_{\mathrm{KL}}(p_{\mathrm{tg}} \Vert p_V) - D_{\mathrm{KL}}(p_{\mathrm{tg}} \Vert p)$, where $D_{\mathrm{KL}}(\cdot \Vert \cdot)$ is the Kullback--Leibler divergence. Most commonly, this is done using an averaged stochastic gradient descent (SGD) algorithm:
\begin{equation}
    \boldsymbol{\alpha}^{(n + 1)} = \boldsymbol{\alpha}^{(n)} - \mu [\nabla \Omega(\Bar{\boldsymbol{\alpha}}^{(n)}) + H_\Omega(\Bar{\boldsymbol{\alpha}}^{(n)}) (\boldsymbol{\alpha}^{(n)} - \Bar{\boldsymbol{\alpha}}^{(n)})].
\end{equation}
Here, $\mu$ is the step size, $\boldsymbol{\alpha}^{(n)}$ the value of the parameter vector $\boldsymbol{\alpha}$ at the $n$-th iteration, and $\Bar{\boldsymbol{\alpha}}^{(n)}$ its cumulative average up to that iteration. For a general functional ansatz, the gradient and Hessian are given by
\begin{equation}
    \frac{\partial \Omega(\boldsymbol{\alpha})}{\partial \alpha_i} = - \Bigl\langle \frac{\partial V(\mathbf{s}; \boldsymbol{\alpha})}{\partial \alpha_i} \Bigr\rangle_{V(\boldsymbol{\alpha})} + \Bigl\langle \frac{\partial V(\mathbf{s}; \boldsymbol{\alpha})}{\partial \alpha_i} \Bigr\rangle_{p_{\mathrm{tg}}},
\end{equation}
\begin{equation}
    \frac{\partial^2 \Omega(\boldsymbol{\alpha})}{\partial \alpha_j \partial \alpha_i} = \operatorname{Cov}\biggl[ \frac{\partial V(\mathbf{s}; \boldsymbol{\alpha})}{\partial \alpha_j}, \frac{\partial V(\mathbf{s}; \boldsymbol{\alpha})}{\partial \alpha_i} \biggr]_{V(\boldsymbol{\alpha})} - \Bigl\langle \frac{\partial^2 V(\mathbf{s}; \boldsymbol{\alpha})}{\partial \alpha_j \partial \alpha_i} \Bigr\rangle_{V(\boldsymbol{\alpha})} + \Bigl\langle \frac{\partial^2 V(\mathbf{s}; \boldsymbol{\alpha})}{\partial \alpha_j \partial \alpha_i} \Bigr\rangle_{p_{\mathrm{tg}}}.
\end{equation}
Here, $\langle \cdots \rangle_{V(\boldsymbol{\alpha})}$ and $\operatorname{Cov}[\cdots]_{V(\boldsymbol{\alpha})}$ denote expectation values and covariances with respect to the biased ensemble associated with the bias potential $V(\boldsymbol{\alpha})$, while $\langle \cdots \rangle_{p_{\mathrm{tg}}}$ denotes expectation values with respect to the target distribution $p_{\mathrm{tg}}(\mathbf{s})$.

It is common to express the potential as a linear combination of basis functions,
\begin{equation}
    V(\mathbf{s}; \boldsymbol{\alpha}) = \sum_{k} \alpha_k f_k(\mathbf{s}).
\end{equation}
In this case, the expressions for the gradient and Hessian simplify to
\begin{equation}
    \frac{\partial \Omega(\boldsymbol{\alpha})}{\partial \alpha_i} = -\bigl\langle f_i(\mathbf{s}) \bigr\rangle_{V(\boldsymbol{\alpha})} + \bigl\langle f_i(\mathbf{s}) \bigr\rangle_{p_{\mathrm{tg}}},
    \label{eq:ves_simplified_gradient}
\end{equation}
\begin{equation}
    \frac{\partial^2 \Omega(\boldsymbol{\alpha})}{\partial \alpha_j \partial \alpha_i} = \operatorname{Cov}\bigl[ f_j(\mathbf{s}), f_i(\mathbf{s}) \bigr]_{V(\boldsymbol{\alpha})}.
    \label{eq:ves_simplified_hessian}
\end{equation}
Here, we use an ansatz that was originally proposed in a different context in \cite{Laio:2015era}:
\begin{equation}
    V(Q; \boldsymbol{\alpha}) = \alpha_1 Q^2 + \alpha_2 \sin^2(Z \pi Q).
    \label{eq:fit_ansatz}
\end{equation}
The first parameter, $\alpha_1 \sim (\chi_t V)^{-1}$ is related to the topological susceptibility, while $\alpha_2$ corresponds to the barrier height and is typically of order $5\text{--}15$ for fine lattices, although this depends strongly on the precise definition of the collective variable. The factor $Z \gtrsim 1$ is a renormalization factor for the topological charge $Q$ and is expected to approach $1$ in the continuum limit. To keep the ansatz linear, we treat $Z$ as a constant; its value can be estimated straightforwardly. This ansatz is intentionally restrictive, since our goal is not to represent arbitrary potentials, but rather to enforce the expected structure of the action barriers and thereby accelerate the buildup.

\begin{figure}[H]
    \centering
    \includegraphics{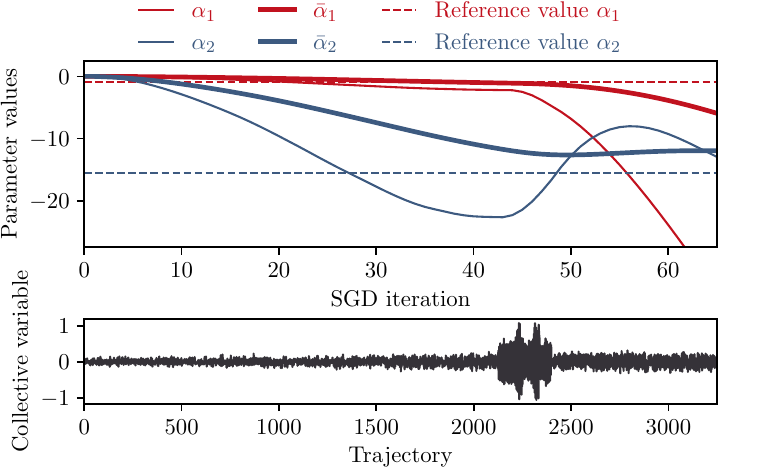}
    \caption{Evolution of the bias parameters (upper panel) and the CV (lower panel) in a VES run for four-dimensional SU(3) gauge theory with the DBW2 action at $L/a = 16$ and $\beta = 1.25$. One SGD iteration corresponds to $50$ HMC trajectories of length $4$. The reference values for $\alpha_1$ and $\alpha_2$ were obtained by fitting \Cref{eq:fit_ansatz} to a bias potential generated in an independent Metadynamics simulation.}
    \label{fig:VES_parameter_evolution}
\end{figure}

\Cref{fig:VES_parameter_evolution} shows the VES parameter evolution and CV time series in a four-dimensional SU(3) simulation with the DBW2 action at $L/a = 16$ and $\beta = 1.25$, using averaged SGD with heavy ball momentum, batches of $50$ HMC trajectories of length $4$, and a flat target distribution.
The simulation becomes unstable after roughly $45$ SGD iterations, with $\alpha_1$ rapidly decreasing to large negative values. This can be attributed to the small batch size relative to the integrated autocorrelation time of the topological charge, which makes the distinction between the barrier terms and the global $Q^2$ term difficult. In our setup, this instability sets in immediately after the first tunneling event, since we use an adaptively expanding integration domain to evaluate the expectation values with respect to the target distribution in \Cref{eq:ves_simplified_gradient}. We expect this behaviour to improve substantially with multiple walkers \cite{doi:10.1021/jp054359r} and by constraining $\alpha_1$ based on physical arguments.

\subsection{Extrapolating the bias potential}
\label{subsec:Extrapolating_the_bias_potential}
\begin{figure}[H]
    \centering
    \includegraphics{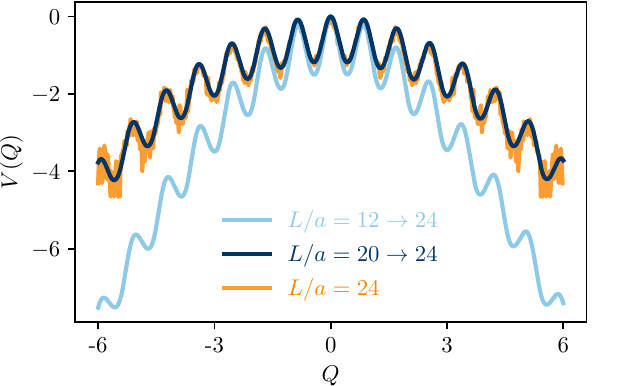}
    \caption{Convolution-based volume extrapolation of bias potentials at $\beta = 6.1912$ in four-dimensional SU(3) gauge theory. The $L/a = 24$ reference potential was reconstructed from an unbiased simulation from \cite{Durr:2025qtq}. The extrapolation from $L/a = 12$ to $L/a = 24$ is inaccurate due to sizeable finite volume effects. Since the volume ratio $24^4/20^4 = 2.0736$ is close to $2$, a good estimate for the bias potential at the larger volume can be obtained by a single convolution.}
    \label{fig:potential_volume_extrapolation}
\end{figure}

A complementary strategy for accelerating the construction of the bias potential is to exploit information obtained at other parameters. Based on the observation that the probability distribution of the sum of independent random variables can be expressed as the convolution of the corresponding individual probability distributions, we can approximate the probability distribution for a large volume simulation by a convolution of probability distributions obtained at smaller volumes and identical parameters. To illustrate the idea, \Cref{fig:potential_volume_extrapolation} compares two extrapolated bias potentials with a reference potential obtained from a simulation at a larger volume.
While the extrapolation from the $L/a = 12$ potential is visibly affected by finite volume effects, the extrapolation from the $L/a = 20$ potential closely matches the reference potential. This directly reflects the finite volume effects of the topological susceptibility; see, for example, Figure~7 in \cite{Durr:2025qtq}.

\section{Improvements to the HMC algorithm}
\label{sec:Improvements_to_the_HMC_algorithm}
In addition to the enhanced sampling techniques discussed in the previous section, we also investigated orthogonal algorithmic improvements that focus on the HMC algorithm itself.

\subsection{Trajectory length tuning}
\label{subsec:Trajectory_length_tuning}
It has been shown in free field theory that the choice of trajectory length can significantly influence simulation efficiency, and that inappropriate choices may even render simulations non-ergodic \cite{Kennedy:2000ju}. Generally, short trajectory lengths tend to produce a suboptimal, diffusive random-walk-like behaviour, whereas excessively long trajectories yield diminishing returns and may eventually become counterproductive due to (approximate) Poincaré recurrences. For related discussions of the impact of trajectory length on autocorrelations in the context of lattice field theory, see \cite{Meyer:2006ty, Schaefer:2010hu, Ostmeyer:2024amv, Yamamoto:2025imx}.

\begin{figure}[H]
    \centering
    \includegraphics{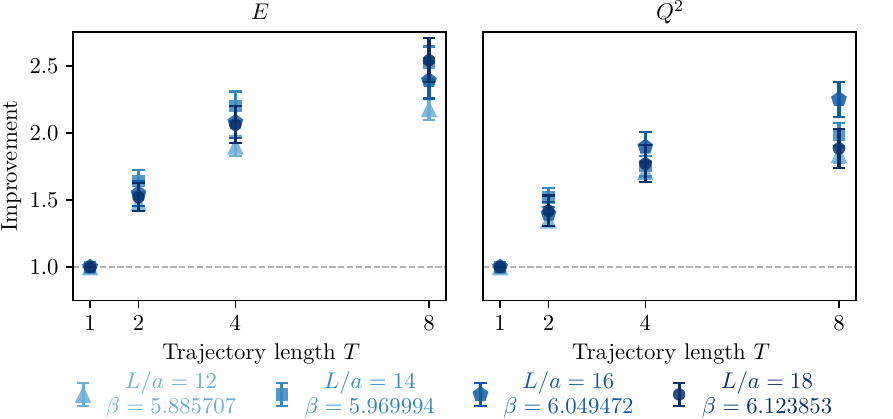}
    \caption{Effect of the HMC trajectory length on the sampling efficiency for the energy density $E$ and the squared topological charge $Q^2$. The additional computational overhead of longer trajectories has already been taken into account.}
    \label{fig:trajectory_length_improvement_e_q2}
\end{figure}

\begin{figure}[H]
    \centering
    \includegraphics{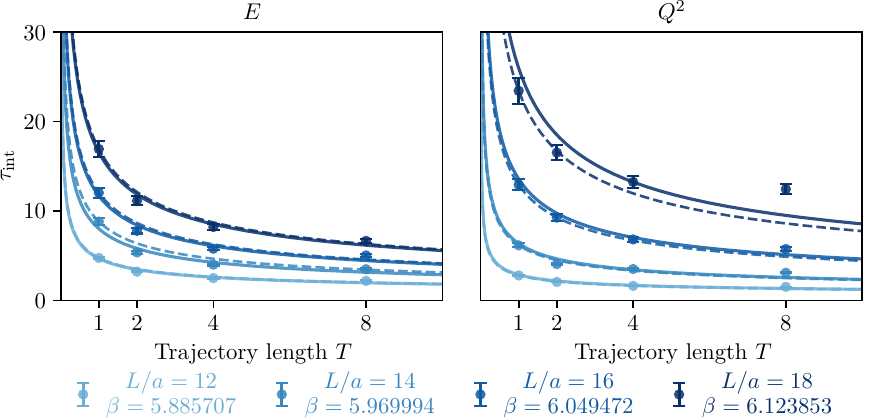}
    \caption{Effect of the HMC trajectory length on the integrated autocorrelation times of the energy density $E$ and the squared topological charge $Q^2$. The additional computational overhead of longer trajectories has already been taken into account. The solid lines are fits of the form $c / \sqrt{T} + 0.5$ to all data points, while the dashed lines correspond to the same functional form with $c = \tau_{\mathrm{int}, T = 1}$.}
    \label{fig:trajectory_length_scaling_e_q2}
\end{figure}

We performed SU(3) simulations with the Wilson action along a line of constant physics, using HMC trajectory lengths $T \in \{1, 2, 4, 8\}$. Measurements were taken every $8$ molecular dynamics units, yielding $10^5$ measurements per run, and integrated autocorrelation times were estimated with the $\Gamma$-method \cite{Wolff:2003sm}. Below, we focus on the energy density $E$ and the squared topological charge $Q^2$, both after 10 steps of stout smearing with smearing parameter $\rho = 0.1$. The resulting gains in sampling efficiency are shown in \Cref{fig:trajectory_length_improvement_e_q2}, and the integrated autocorrelation times in \Cref{fig:trajectory_length_scaling_e_q2}.

For both observables, increasing the trajectory length beyond $T = 1$ leads to clear reductions in the cost-normalized integrated autocorrelation times. The same qualitative behaviour can be observed for other observables, including the action, larger smeared Wilson loops, and Polyakov loops. However, we want to emphasize that the optimal trajectory length is generally observable-dependent, since it depends on how a given observable couples to the modes of the Markov chain.

There is no indication that the dynamical critical exponent is reduced; rather, the integrated autocorrelation times appear to be reduced by an approximately constant factor that is independent of the lattice spacing. Similar improvements seem to persist in simulations with a bias potential and in simulations including dynamical fermions.

The observed trend is roughly compatible with a $1 / \sqrt{T}$ improvement, as already suggested in \cite{Schaefer:2010hu}. For the energy density, this description seems to work reasonably well over the explored range. For $Q^2$, however, the agreement is less convincing, which may indicate that the topological slow modes behave differently at a fundamental level.

These observations may motivate the use of more sophisticated HMC variants in which the trajectory length is not fixed a priori. Examples include the No-U-Turn Sampler (NUTS) \cite{hoffman2014no} and the Exhaustive HMC (XHMC) \cite{betancourt2016}, which integrate the equations of motion until some termination criterion is met. Their application to lattice gauge theories would likely require a gauge-invariant termination criterion. It would also be interesting to define the termination criterion in terms of collective variables, i.e., to evolve the trajectory until a U-turn in CV space occurs.

\subsection{Recycling HMC}
\label{subsec:Recycling_HMC}
Usually, only the configurations at the endpoints of HMC trajectories are used, because it is not immediately clear how to incorporate intermediate configurations given the final accept--reject step at the end of the trajectory. As shown in \cite{Nishimura_2020}, these intermediate configurations can nevertheless be used to construct improved, unbiased estimators by performing additional accept--reject steps against the initial configuration for each intermediate configuration. The actual Markov chain remains unchanged compared to the conventional HMC.

There is an obvious synergy with the previously discussed use of increased trajectory lengths, since one can still perform the same number of measurements per unit of molecular dynamics time as with shorter trajectories. The resulting improvements can be substantial for the buildup of the bias potential and for observables with short autocorrelation times, $\tau_{\mathrm{int}} \lesssim T$. The computational overhead from the additional accept--reject steps depends on the simulation setup, and can be negligible if the fermionic forces are already computed to full precision during the trajectory. Since we are primarily interested in obtaining a good approximation to the bias potential, and since we observed acceptance rates around \SI{99}{\percent} in our simulations, we chose to omit these additional accept--reject steps in our experiments. Combined with longer trajectory lengths, this procedure accelerates the buildup of our bias potentials by approximately an order of magnitude.

\subsection{Repelling-Attracting HMC}
\label{subsec:Repelling-Attracting_HMC}
The Repelling-Attracting HMC (RAHMC) algorithm \cite{vishwanath2024} is a modification of the conventional HMC algorithm designed to improve sampling efficiency for multimodal distributions. To accomplish this, an additional friction parameter $\gamma > 0$ is introduced, and the integration of the classical equations of motion is split into two stages: An initial mode-repelling phase, in which the momenta are updated according to
\begin{equation}
    P_{\mu}(n) \to P'_{\mu}(n) = e^{\gamma \epsilon} P_{\mu}(n) - \epsilon \frac{\partial S(U)}{\partial U_{\mu}(n)},
\end{equation}
followed by a mode-attracting phase, in which the momenta are updated according to
\begin{equation}
    P_{\mu}(n) \to P'_{\mu}(n) = e^{-\gamma \epsilon} \biggl(P_{\mu}(n) - \epsilon \frac{\partial S(U)}{\partial U_{\mu}(n)} \biggr).
\end{equation}
Here, $\epsilon$ denotes the step size of the integration step. Although the original work proved that this algorithm is both volume-preserving and reversible, it established only an upper bound on the energy drift. In our experiments in SU(3) gauge theory, we found the energy violations to be unacceptably large over the parameter range we explored, with friction parameters $\gamma \in [0.005, 0.1]$ and trajectory lengths $T \in [0.5, 8]$. Moreover, the acceptance rates deteriorated rapidly with increasing numbers of degrees of freedom, even for comparatively small integrator step sizes. This appears to rule out the application of this algorithm in its current form in the context of lattice QCD. It is worth noting, however, that the original authors used a dual-averaging procedure to tune the step size, trajectory length, and friction parameter automatically, whereas we explored the parameter space manually.
\section{Conclusion}
\label{sec:Conclusion}
At present, the most successful strategy for constructing bias potentials---and the one used in our production runs---combines longer HMC trajectories ($T = 4$ or $T = 8$) with Recycling HMC, i.e., the use of intermediate configurations along the trajectory as described in \Cref{subsec:Recycling_HMC}. This accelerates the buildup of the bias potential by around an order of magnitude. Once a bias potential has been obtained, we extrapolate it to larger volumes following the strategy discussed in \Cref{subsec:Extrapolating_the_bias_potential} to obtain a reasonable initial guess that can then be refined in a short simulation at the target volume.

The other methods discussed here have not yet led to any improvements in our present setup. Nevertheless, variationally enhanced sampling remains a promising direction for future work, since a parametric description of the bias potential could make it possible to extrapolate the potential not only in volume, but also across other simulation parameters.

\acknowledgments
We gratefully acknowledge helpful discussions with Szabolcs Borsanyi, Stephan Dürr, Jacob Finkenrath, Anthony Kennedy, Piyush Kumar, and Kalman Szabo. T.E. is supported by the Deutsche Forschungsgemeinschaft (DFG, grant No. HO 4177/1-1).

\bibliographystyle{JHEP}
\setlength{\bibsep}{0pt plus 0.9ex}
\bibliography{literature.bib}

\providecommand{\href}[2]{#2}\begingroup\raggedright\begin{thebibliography}{10}

\bibitem{Alles:1996vn}
B.~Alles, G.~Boyd, M.~D'Elia, A.~Di~Giacomo and E.~Vicari, \emph{{Hybrid Monte Carlo and topological modes of full QCD}}, \href{https://doi.org/10.1016/S0370-2693(96)01247-6}{\emph{Phys. Lett. B} {\bfseries 389} (1996) 107} [\href{https://arxiv.org/abs/hep-lat/9607049}{{\ttfamily hep-lat/9607049}}].

\bibitem{Orginos:2001xa}
{\scshape RBC Collaboration} collaboration, \emph{{Chiral properties of domain wall fermions with improved gauge actions}}, \href{https://doi.org/10.1016/S0920-5632(01)01827-8}{\emph{Nucl. Phys. B Proc. Suppl.} {\bfseries 106} (2002) 721} [\href{https://arxiv.org/abs/hep-lat/0110074}{{\ttfamily hep-lat/0110074}}].

\bibitem{DeGrand:2002vu}
T.A.~DeGrand, A.~Hasenfratz and T.G.~Kovacs, \emph{{Improving the chiral properties of lattice fermions}}, \href{https://doi.org/10.1103/PhysRevD.67.054501}{\emph{Phys. Rev. D} {\bfseries 67} (2003) 054501} [\href{https://arxiv.org/abs/hep-lat/0211006}{{\ttfamily hep-lat/0211006}}].

\bibitem{Noaki:2002ai}
{\scshape RBC Collaboration} collaboration, \emph{{Calculation of weak matrix elements in domain wall QCD with the DBW2 gauge action}}, \href{https://doi.org/10.1016/S0920-5632(03)01552-4}{\emph{Nucl. Phys. B Proc. Suppl.} {\bfseries 119} (2003) 362} [\href{https://arxiv.org/abs/hep-lat/0211013}{{\ttfamily hep-lat/0211013}}].

\bibitem{Aoki:2005ga}
Y.~Aoki et~al., \emph{{The Kaon B-parameter from quenched domain-wall QCD}}, \href{https://doi.org/10.1103/PhysRevD.73.094507}{\emph{Phys. Rev. D} {\bfseries 73} (2006) 094507} [\href{https://arxiv.org/abs/hep-lat/0508011}{{\ttfamily hep-lat/0508011}}].

\bibitem{Schaefer:2009xx}
S.~Schaefer, R.~Sommer and F.~Virotta, \emph{{Investigating the critical slowing down of QCD simulations}}, \href{https://doi.org/10.22323/1.091.0032}{\emph{PoS} {\bfseries LAT2009} (2010) 032} [\href{https://arxiv.org/abs/0910.1465}{{\ttfamily 0910.1465}}].

\bibitem{Schaefer:2010hu}
{\scshape ALPHA Collaboration} collaboration, \emph{{Critical slowing down and error analysis in lattice QCD simulations}}, \href{https://doi.org/10.1016/j.nuclphysb.2010.11.020}{\emph{Nucl. Phys. B} {\bfseries 845} (2011) 93} [\href{https://arxiv.org/abs/1009.5228}{{\ttfamily 1009.5228}}].

\bibitem{Kanwar:2025wuc}
G.~Kanwar and O.~Vega, \emph{{Spectral Diffusion for Sampling on ${\rm SU}(N)$}},  \href{https://arxiv.org/abs/2512.19877}{{\ttfamily 2512.19877}}.

\bibitem{Bonanno:2026tle}
C.~Bonanno, A.~Bulgarelli, E.~Cellini, A.~Nada, D.~Panfalone, D.~Vadacchino et~al., \emph{{A scalable flow-based approach to mitigate topological freezing}},  \href{https://arxiv.org/abs/2601.20708}{{\ttfamily 2601.20708}}.

\bibitem{Matsumoto:2026puv}
A.~Matsumoto, M.~Hirasawa, J.~Nishimura and A.~Yosprakob, \emph{{Phase diagram of 4D SU(3) Yang-Mills theory at $\theta=\pi$ via imaginary theta simulations}},  \href{https://arxiv.org/abs/2603.09604}{{\ttfamily 2603.09604}}.

\bibitem{Singh:2026zls}
S.~Singh and L.~Funcke, \emph{{Normalizing-flow-based density of states for (1+1)D U(1) lattice gauge theory with a $\theta$-term}},  \href{https://arxiv.org/abs/2603.12501}{{\ttfamily 2603.12501}}.

\bibitem{Finkenrath:2023sjg}
J.~Finkenrath, \emph{{Review on Algorithms for dynamical fermions}}, \href{https://doi.org/10.22323/1.430.0227}{\emph{PoS} {\bfseries LATTICE2022} (2023) 227} [\href{https://arxiv.org/abs/2402.11704}{{\ttfamily 2402.11704}}].

\bibitem{Kanwar:2024ujc}
G.~Kanwar, \emph{{Flow-based sampling for lattice field theories}}, \href{https://doi.org/10.22323/1.453.0114}{\emph{PoS} {\bfseries LATTICE2023} (2024) 114} [\href{https://arxiv.org/abs/2401.01297}{{\ttfamily 2401.01297}}].

\bibitem{Boyle:2024nlh}
P.A.~Boyle, \emph{{Advances in algorithms for solvers and gauge generation}}, \href{https://doi.org/10.22323/1.453.0122}{\emph{PoS} {\bfseries LATTICE2023} (2024) 122} [\href{https://arxiv.org/abs/2401.16620}{{\ttfamily 2401.16620}}].

\bibitem{Finkenrath:2024ptc}
J.~Finkenrath, \emph{{Future trends in lattice QCD simulations}}, \href{https://doi.org/10.22323/1.451.0009}{\emph{PoS} {\bfseries EuroPLEx2023} (2024) 009}.

\bibitem{Luscher:2009eq}
M.~Luscher, \emph{{Trivializing maps, the Wilson flow and the HMC algorithm}}, \href{https://doi.org/10.1007/s00220-009-0953-7}{\emph{Commun. Math. Phys.} {\bfseries 293} (2010) 899} [\href{https://arxiv.org/abs/0907.5491}{{\ttfamily 0907.5491}}].

\bibitem{Albergo:2019eim}
M.S.~Albergo, G.~Kanwar and P.E.~Shanahan, \emph{{Flow-based generative models for Markov chain Monte Carlo in lattice field theory}}, \href{https://doi.org/10.1103/PhysRevD.100.034515}{\emph{Phys. Rev. D} {\bfseries 100} (2019) 034515} [\href{https://arxiv.org/abs/1904.12072}{{\ttfamily 1904.12072}}].

\bibitem{Wang:2023exq}
L.~Wang, G.~Aarts and K.~Zhou, \emph{{Diffusion models as stochastic quantization in lattice field theory}}, \href{https://doi.org/10.1007/JHEP05(2024)060}{\emph{JHEP} {\bfseries 05} (2024) 060} [\href{https://arxiv.org/abs/2309.17082}{{\ttfamily 2309.17082}}].

\bibitem{Hasenbusch:2017unr}
M.~Hasenbusch, \emph{{Fighting topological freezing in the two-dimensional $CP^{N-1}$ model}}, \href{https://doi.org/10.1103/PhysRevD.96.054504}{\emph{Phys. Rev. D} {\bfseries 96} (2017) 054504} [\href{https://arxiv.org/abs/1706.04443}{{\ttfamily 1706.04443}}].

\bibitem{Bonanno:2024zyn}
C.~Bonanno, G.~Clemente, M.~D'Elia, L.~Maio and L.~Parente, \emph{{Full QCD with milder topological freezing}}, \href{https://doi.org/10.1007/JHEP08(2024)236}{\emph{JHEP} {\bfseries 08} (2024) 236} [\href{https://arxiv.org/abs/2404.14151}{{\ttfamily 2404.14151}}].

\bibitem{Eichhorn:2023uge}
T.~Eichhorn, G.~Fuwa, C.~Hoelbling and L.~Varnhorst, \emph{{Parallel tempered metadynamics: Overcoming potential barriers without surfing or tunneling}}, \href{https://doi.org/10.1103/PhysRevD.109.114504}{\emph{Phys. Rev. D} {\bfseries 109} (2024) 114504} [\href{https://arxiv.org/abs/2307.04742}{{\ttfamily 2307.04742}}].

\bibitem{Eichhorn:2025aph}
T.~Eichhorn, G.~Fuwa, C.~Hoelbling and L.~Varnhorst, \emph{{Parallel Tempered Metadynamics}}, \href{https://doi.org/10.22323/1.466.0061}{\emph{PoS} {\bfseries LATTICE2024} (2025) 61} [\href{https://arxiv.org/abs/2503.09747}{{\ttfamily 2503.09747}}].

\bibitem{Duane:1987de}
S.~Duane, A.D.~Kennedy, B.J.~Pendleton and D.~Roweth, \emph{{Hybrid Monte Carlo}}, \href{https://doi.org/10.1016/0370-2693(87)91197-X}{\emph{Phys. Lett. B} {\bfseries 195} (1987) 216}.

\bibitem{Morningstar:2003gk}
C.~Morningstar and M.J.~Peardon, \emph{{Analytic smearing of SU(3) link variables in lattice QCD}}, \href{https://doi.org/10.1103/PhysRevD.69.054501}{\emph{Phys. Rev. D} {\bfseries 69} (2004) 054501} [\href{https://arxiv.org/abs/hep-lat/0311018}{{\ttfamily hep-lat/0311018}}].

\bibitem{GarciaPerez:1993lic}
M.~Garcia~Perez, A.~Gonzalez-Arroyo, J.R.~Snippe and P.~van Baal, \emph{{Instantons from over - improved cooling}}, \href{https://doi.org/10.1016/0550-3213(94)90631-9}{\emph{Nucl. Phys. B} {\bfseries 413} (1994) 535} [\href{https://arxiv.org/abs/hep-lat/9309009}{{\ttfamily hep-lat/9309009}}].

\bibitem{deForcrand:1997esx}
P.~de~Forcrand, M.~Garcia~Perez and I.-O.~Stamatescu, \emph{{Topology of the SU(2) vacuum: A Lattice study using improved cooling}}, \href{https://doi.org/10.1016/S0550-3213(97)00275-7}{\emph{Nucl. Phys. B} {\bfseries 499} (1997) 409} [\href{https://arxiv.org/abs/hep-lat/9701012}{{\ttfamily hep-lat/9701012}}].

\bibitem{Tanizaki:2024zsu}
Y.~Tanizaki, A.~Tomiya and H.~Watanabe, \emph{{Lattice gradient flows (de-)stabilizing topological sectors}}, \href{https://doi.org/10.1007/JHEP04(2025)123}{\emph{JHEP} {\bfseries 04} (2025) 123} [\href{https://arxiv.org/abs/2411.14812}{{\ttfamily 2411.14812}}].

\bibitem{Butti:2025rnu}
P.~Butti, M.~Della~Morte, B.~J{\"a}ger, S.~Martins and J.T.~Tsang, \emph{{Comparison of smoothening flows for the topological charge in QCD-like theories}}, \href{https://doi.org/10.1103/53vh-wm6v}{\emph{Phys. Rev. D} {\bfseries 112} (2025) 014504} [\href{https://arxiv.org/abs/2504.10197}{{\ttfamily 2504.10197}}].

\bibitem{Laio_2002}
A.~Laio and M.~Parrinello, \emph{Escaping free-energy minima}, \href{https://doi.org/10.1073/pnas.202427399}{\emph{Proc. Natl. Acad. Sci.} {\bfseries 99} (2002) 12562}.

\bibitem{PhysRevLett.113.090601}
O.~Valsson and M.~Parrinello, \emph{Variational approach to enhanced sampling and free energy calculations}, \href{https://doi.org/10.1103/PhysRevLett.113.090601}{\emph{Phys. Rev. Lett.} {\bfseries 113} (2014) 090601}.

\bibitem{Laio:2015era}
A.~Laio, G.~Martinelli and F.~Sanfilippo, \emph{{Metadynamics surfing on topology barriers: the $CP^{N-1}$ case}}, \href{https://doi.org/10.1007/JHEP07(2016)089}{\emph{JHEP} {\bfseries 07} (2016) 089} [\href{https://arxiv.org/abs/1508.07270}{{\ttfamily 1508.07270}}].

\bibitem{doi:10.1021/jp054359r}
P.~Raiteri, A.~Laio, F.L.~Gervasio, C.~Micheletti and M.~Parrinello, \emph{Efficient reconstruction of complex free energy landscapes by multiple walkers metadynamics}, \href{https://doi.org/10.1021/jp054359r}{\emph{J. Phys. Chem. B} {\bfseries 110} (2006) 3533}.

\bibitem{Durr:2025qtq}
S.~Durr and G.~Fuwa, \emph{{Topological susceptibility and excess kurtosis in SU(3) Yang-Mills theory}}, \href{https://doi.org/10.1103/dw3m-vcdm}{\emph{Phys. Rev. D} {\bfseries 113} (2026) 054508} [\href{https://arxiv.org/abs/2501.08217}{{\ttfamily 2501.08217}}].

\bibitem{Kennedy:2000ju}
A.D.~Kennedy and B.~Pendleton, \emph{{Cost of the generalized hybrid Monte Carlo algorithm for free field theory}}, \href{https://doi.org/10.1016/S0550-3213(01)00129-8}{\emph{Nucl. Phys. B} {\bfseries 607} (2001) 456} [\href{https://arxiv.org/abs/hep-lat/0008020}{{\ttfamily hep-lat/0008020}}].

\bibitem{Meyer:2006ty}
H.B.~Meyer, H.~Simma, R.~Sommer, M.~Della~Morte, O.~Witzel and U.~Wolff, \emph{{Exploring the HMC trajectory-length dependence of autocorrelation times in lattice QCD}}, \href{https://doi.org/10.1016/j.cpc.2006.08.002}{\emph{Comput. Phys. Commun.} {\bfseries 176} (2007) 91} [\href{https://arxiv.org/abs/hep-lat/0606004}{{\ttfamily hep-lat/0606004}}].

\bibitem{Ostmeyer:2024amv}
J.~Ostmeyer and P.~Buividovich, \emph{{Minimal Autocorrelation in Hybrid Monte Carlo simulations using Exact Fourier Acceleration}},  \href{https://arxiv.org/abs/2404.09723}{{\ttfamily 2404.09723}}.

\bibitem{Yamamoto:2025imx}
S.~Yamamoto, P.~Boyle, T.~Izubuchi, L.~Jin, C.~Lehner and N.~Matsumoto, \emph{{Improvement in Autocorrelation Times Measured by the Master-Field Technique using Field Transformation HMC in 2+1 Domain Wall Fermion Simulations}}, \href{https://doi.org/10.22323/1.466.0034}{\emph{PoS} {\bfseries LATTICE2024} (2025) 034} [\href{https://arxiv.org/abs/2502.05452}{{\ttfamily 2502.05452}}].

\bibitem{Wolff:2003sm}
{\scshape ALPHA Collaboration} collaboration, \emph{{Monte Carlo errors with less errors}}, \href{https://doi.org/10.1016/S0010-4655(03)00467-3}{\emph{Comput. Phys. Commun.} {\bfseries 156} (2004) 143} [\href{https://arxiv.org/abs/hep-lat/0306017}{{\ttfamily hep-lat/0306017}}].

\bibitem{hoffman2014no}
M.D.~Hoffman, A.~Gelman et~al., \emph{The no-u-turn sampler: adaptively setting path lengths in hamiltonian monte carlo.}, {\emph{J. Mach. Learn. Res.} {\bfseries 15} (2014) 1593}.

\bibitem{betancourt2016}
M.~Betancourt, \emph{Identifying the optimal integration time in hamiltonian monte carlo},  \href{https://arxiv.org/abs/1601.00225}{{\ttfamily 1601.00225}}.

\bibitem{Nishimura_2020}
A.~Nishimura and D.~Dunson, \emph{Recycling intermediate steps to improve hamiltonian monte carlo}, \href{https://doi.org/10.1214/19-ba1171}{\emph{Bayesian Analysis} {\bfseries 15} (2020) }.

\bibitem{vishwanath2024}
S.~Vishwanath and H.~Tak, \emph{Repelling-attracting hamiltonian monte carlo},  \href{https://arxiv.org/abs/2403.04607}{{\ttfamily 2403.04607}}.

\end{thebibliography}\endgroup
\end{document}